# Beyond Low Earth Orbit: Biomonitoring, Artificial Intelligence, and Precision Space Health

Ryan T. Scott[1], Erik L. Antonsen[2], Lauren M. Sanders[3], Jaden J.A. Hastings[4], Seung-min Park[5], Graham Mackintosh[6], Robert J. Reynolds[7], Adrienne L. Hoarfrost[8], Aenor Sawyer[9], Casey S. Greene[10], Benjamin S. Glicksberg[11], Corey A. Theriot[12,13], Daniel C. Berrios[1], Jack Miller[1], Joel Babdor[14], Richard Barker[15], Sergio E. Baranzini[16], Afshin Beheshti[1], Stuart Chalk[17], Guillermo M. Delgado-Aparicio[18], Melissa Haendel[19], Arif A. Hamid[20], Philip Heller[21], Daniel Jamieson[22], Katelyn J. Jarvis[9], John Kalantari[23], Kia Khezeli[23], Svetlana V. Komarova[24], Matthieu Komorowski[25], Prachi Kothiyal[26], Ashish Mahabal[27], Uri Manor[28], Hector Garcia Martin[29,30,31], Christopher E. Mason[4], Mona Matar[32], George I. Mias[33], Jerry G. Myers, Jr.[32], Charlotte Nelson[16], Jonathan Oribello[3], Patricia Parsons-Wingerter[34], R. K. Prabhu[35], Amina Ann Qutub[36], Jon Rask[37], Amanda Saravia-Butler[38], Suchi Saria[39,40], Nitin Kumar Singh[41], Frank Soboczenski[42], Michael Snyder[43], Karthik Soman[16], David Van Valen[44], Kasthuri Venkateswaran[41], Liz Warren[45], Liz Worthey[46], Jason H. Yang[47], Marinka Zitnik[48], Sylvain V. Costes[49+]

[1] KBR, Space Biosciences Division, NASA Ames Research Center, Moffett Field, CA 94035, USA.
[2] Department of Emergency Medicine, Center for Space Medicine, Baylor College of Medicine, Houston, TX 77030, USA.
[3] Blue Marble Space Institute of Science, Space Biosciences Division, NASA Ames Research Center, Moffett Field, CA 94035, USA.
[4] Department of Physiology and Biophysics, Weill Cornell Medicine, New York, NY 10065, USA.
[5] Department of Urology, Department of Radiology, Stanford University School of Medicine, Stanford, CA 94305, USA.
[6] Bay Area Environmental Research Institute, NASA Ames Research Center, Moffett Field, CA 94035, USA.
[7] Mortality Research & Consulting, Inc., Houston, TX 77058, USA.
[8] Universities Space Research Association (USRA), Space Biosciences Division, NASA Ames Research Center, Moffett Field, CA 94035, USA.
[9] UC Space Health, Department of Orthopaedic Surgery, University of California, San Francisco, San Francisco, CA 94143, USA.
[10] Center for Health AI, Department of Biochemistry and Molecular Genetics, University of Colorado School of Medicine, Anschutz Medical Campus, Aurora, CO 80045, USA.
[11] Hasso Plattner Institute for Digital Health at Mount Sinai, Department of Genetics and Genomic Sciences, Icahn School of Medicine at Mount Sinai, New York, NY 10065, USA.
[12] Department of Preventive Medicine and Community Health, UTMB, Galveston, TX 77551 USA.
[13] Human Health and Performance Directorate, NASA Johnson Space Center, Houston, TX 77058, USA.
[14] Department of Microbiology and Immunology, Department of Otolaryngology, Head and Neck Surgery, University of California San Francisco, San Francisco, CA 94143, USA.
[15] The Gilroy AstroBiology Research Group, The University of Wisconsin, Madison, Madison, WI 53706, USA.
[16] Weill Institute for Neurosciences, Department of Neurology, University of California San Francisco, San Francisco, CA 94158, USA.
[17] Department of Chemistry, University of North Florida, Jacksonville, FL 32224, USA.
[18] Data Science Analytics, Georgia Institute of Technology, Lima 15024, Peru.
[19] Center for Health AI, University of Colorado School of Medicine, Anschutz Medical Campus, Aurora, CO 80238, USA.
[20] Department of Neuroscience, University of Minnesota, Minneapolis, MN 55455, USA.

[21] Department of Computer Science, College of Science, San José State University, San Jose, CA 95192, USA.
[22] Biorelate, Manchester, M15 6SE, United Kingdom.
[23] Center for Individualized Medicine, Department of Surgery, Department of Quantitative Health Sciences, Mayo Clinic, Rochester, MN 55905, USA.
[24] Faculty of Dental Medicine and Oral Health Sciences, McGill University, Montreal, Quebec, H4A 0A9, Canada.
[25] Faculty of Medicine, Department of Surgery and Cancer, Imperial College London, London, SW7 2AZ, United Kingdom.
[26] SymbioSeq LLC, NASA Johnson Space Center, Ashburn, VA 20148, USA.
[27] Center for Data Driven Discovery, California Institute of Technology, Pasadena, CA 91125, USA.
[28] Waitt Advanced Biophotonics Center, Chan-Zuckerberg Imaging Scientist Fellow, Salk Institute for Biological Studies, La Jolla, CA 92037, USA.
[29] Biological Systems and Engineering Division, Lawrence Berkeley National Lab, Berkeley, CA 94608, USA.
[30] DOE Agile BioFoundry, Emeryville, CA 94608, USA.
[31] Joint BioEnergy Institute, Emeryville, CA 94608, USA.
[32] Human Research Program Cross-cutting Computational Modeling Project, NASA John H. Glenn Research Center, Cleveland, OH 44135, USA.
[33] Institute for Quantitative Health Science and Engineering, Department of Biochemistry and Molecular Biology, Michigan State University, East Lansing, MI 48824, USA.
[34] Low Exploration Gravity Technology, NASA John H. Glenn Research Center, Cleveland, OH 44135, USA.
[35] Universities Space Research Association (USRA), Human Research Program Cross-cutting Computational Modeling Project, NASA John H. Glenn Research Center, Cleveland, OH 44135, USA.
[36] AI MATRIX Consortium, Department of Biomedical Engineering, University of Texas, San Antonio and UT Health Sciences, San Antonio, TX 78249, USA.
[37] Office of the Center Director, NASA Ames Research Center, Moffett Field, CA 94035, USA.
[38] Logyx, Space Biosciences Division, NASA Ames Research Center, Moffett Field, CA 94035, USA.
[39] Computer Science, Statistics, and Health Policy, Johns Hopkins University, Baltimore, MD 21218, USA.
[40] ML, AI and Healthcare Lab, Bayesian Health, New York, NY 21202, USA.
[41] Biotechnology and Planetary Protection Group, Jet Propulsion Laboratory, Pasadena, CA 91106, USA.
[42] SPHES, Medical Faculty, King's College London, London, WC2R 2LS, United Kingdom.
[43] Department of Genetics, Stanford School of Medicine, Stanford, CA 94305 USA.
[44] Department of Biology, California Institute of Technology, Pasadena, CA 91125, USA.
[45] ISS National Laboratory, Center for the Advancement of Science in Space, Melbourne, FL 32940, USA.
[46] UAB Center for Computational Biology and Data Science, University of Alabama, Birmingham, Birmingham, AL 35223, USA.
[47] Center for Emerging and Re-Emerging Pathogens, Department of Microbiology, Biochemistry and Molecular Genetics, Rutgers New Jersey Medical School, Newark, NJ 07103, USA.
[48] Department of Biomedical Informatics, Harvard Medical School, Harvard Data Science, Broad Institute of MIT and Harvard, Harvard University, Boston, MA 02115, USA.
[49] Space Biosciences Division, NASA Ames Research Center, Moffett Field, CA 94035, USA.

[+]Corresponding Author Information:
Sylvain V. Costes <sylvain.v.costes@nasa.gov>

**Abstract**

Human space exploration beyond low Earth orbit will involve missions of significant distance and duration. To effectively mitigate myriad space health hazards, paradigm shifts in data and space health systems are necessary to enable Earth-independence, rather than Earth-reliance. Promising developments in the fields of artificial intelligence and machine learning for biology and health can address these needs. We propose an appropriately autonomous and intelligent Precision Space Health system that will monitor, aggregate, and assess biomedical statuses; analyze and predict personalized adverse health outcomes; adapt and respond to newly accumulated data; and provide preventive, actionable, and timely insights to individual deep space crew members and iterative decision support to their crew medical officer. Here we present a summary of recommendations from a workshop organized by the National Aeronautics and Space Administration, on future applications of artificial intelligence in space biology and health. In the next decade, biomonitoring technology, biomarker science, spacecraft hardware, intelligent software, and streamlined data management must mature and be woven together into a Precision Space Health system to enable humanity to thrive in deep space.

**Introduction**

Astronauts face hazards unique to spaceflight such as ionizing radiation, altered gravitational fields, accelerated day-night cycles, confined isolation, hostile-closed environments, distance-duration from Earth[1], planetary dust-regolith[2], and extreme temperatures/atmospheres[3,4]. As astronauts experience these hazards, the body responds by adapting and deconditioning over the duration of the exposure with the potential for synergistic effects as the exposures persist[1,5]. As humanity plans exploration of deep space and planetary-class missions (i.e., cis-Lunar, Mars), astronauts would benefit from access to optimally scoped healthcare and medical systems to ensure success for a given mission. However, many current gaps in knowledge and technological challenges exist which limit the ability to accurately predict the necessary surveillance and mitigation capabilities needed for deep space missions.

Human spaceflight has predominantly been conducted in low Earth orbit (LEO), with access to substantial real-time support from Mission Control Center (MCC) flight surgeons and engineers. In contrast, deep-space crews will be confronted with (1) high-latency communications that prohibit real-time support[6,7], (2) data bandwidth and power constraints[6,8], (3) infrequent resupply[6,7], (4) carrying only essential and effective medications which may degrade over time[9], (5) an inability to evacuate or be quickly rescued[6,7], and (6) greater exposure to solar and galactic cosmic radiation[8,10–12].

In LEO, the current paradigm for data management, data acquisition-monitoring technologies, and medical decision making has been defined by its proximity to Earth, with no urgent need for autonomy from MCC. Biological and health data acquisition systems in LEO have revolved around biosensors and monitoring devices often dedicated to specific experiments and batteries of health tests. Data have been nestled within specific pipelines due to privacy and logistical constraints resulting in limited or restricted access. In addition, these biomedical systems are not fully integrated into any *in situ* analytics or real-time on-board reporting[13–15]. The current LEO medical planning model focuses on estimating the likelihood of specific medical conditions[16] and provides training for onboard resources to react with in-flight diagnostic and therapeutic capabilities. There are only episodic in-flight assessments, with limited or no onboard data analytics for usage toward crew-centered decision making. Weekly private medical conferences and private psychological conferences are performed in real-time with MCC through audio-video streaming[17]. Of note, there have been initial on-board demonstration projects to begin establishing the capability of the crew for autonomous medical activities[18], and sequencing of microbes that could be relevant for human health[19].

The transition from LEO to deep space missions will present novel operational health requirements. To develop a new paradigm for deep space travel, where real-time communication, massive data transfer, and immediate technical guidance is not an option, the development of multi-layered, efficient, and automated biomedical monitoring, predictive analytics, and clinical decision support must all be integrated into an intelligent deep space health system[13,20]. Redistribution of medical decisions, biological data, and data management responsibilities must occur among the participants in this process (i.e., the on-board Crew Health and Performance [CHP] system, the crew themselves, the crew medical officer [CMO], the Environmental Control and Life Support system [ECLS], MCC, and others[21–24]). Such a system must be maximally autonomous when appropriate for the mission and system, increase crew autonomy from Earth, and avail crew of precious time. The system must be as autonomous as possible, but with an essential link to human crew members to assess, evaluate, and act on the resulting data. It will need to identify, predict and provide health solutions to problems before they arise, as well as continuously managing and analyzing an expanding accumulation of environmental, biological and health data. It should provide the CMO with explainable insights into the complexities between the biological environment, crew health, and mission-specific requirements (e.g., destination, duration, vehicle, habitat), as well as offer predictive outcomes depending on courses of treatment (or no treatment).

There is an opportunity for spaceflight biology, health, and medicine to leverage the advances in computer science for deep space. Artificial intelligence (AI), machine learning (ML), and biological-computational modeling all utilize sample data to create a representation of a system that can predict an outcome of interest on future, previously unseen data[25]. We envision an AI-driven, proactive 'Precision Space Health' system, ensuring that care is predictive, preventative, participatory, and personalized[26,27]. Such an approach supports the crew themselves and their CMO to make evidence-based health decisions.

This article presents a decadal vision and summarizes the content from a workshop organized by the National Aeronautics and Space Administration (NASA) entitled 'Workshop on Artificial Intelligence and Modeling for Space Biology.' A parallel article from the workshop reviews AI and ML challenges and opportunities for fundamental space biological research (Sanders et al., 2021 [unpublished preprint]; Supplements 1 & 2). In this review, we summarize current AI, modeling, and ML methods that we believe could be deployed to address deep space health challenges. Of note, this effort toward actualizing a 'Precision Space Health' system, simultaneously advances the emergence of AI in healthcare on Earth[28–30].

**Precision Space Health**

Precision Medicine (PM) refers to personalized medical treatment tailored to the individual characteristics of each patient[31], including genetic predisposition, behavioral influences (exercise, nutrition, stress) and environmental influences (physical and social determinants). Both medical history and current health data are required to enable accurate and effective health insight. Leveraging PM principles, Precision Health (PH) emphasizes disease prevention and early detection via individualized, longitudinal monitoring[32]. Thus, a PM/PH approach is uniquely well-suited for deep space missions[33]. Indeed, in 2014, the National Academy of Sciences recommended an ethical framework for long duration and exploration spaceflight that includes the responsibility to have health standards continually evolve, improve, and be informed by data[34].

Although PM/PH frameworks are still in early development in terrestrial clinical practice, their principles and framework hold great potential for long duration and distant spaceflight. We propose the necessity for a Precision Space Health system that integrates longitudinal clinical, biomarker, human 'omics, behavioral, and microbiome data about an individual in a healthy state in order to facilitate automated and early detection of pathogenic changes (**Figure 1**)[35–42]. With a small number of astronauts, NASA has already invested significant resources in baselining health information at selection and surveils health throughout their career and beyond. An excellent proof-of-concept of longitudinal collection of multiple data types from astronauts was seen in the 'Twins Study', the first of its kind to characterize two genetically identical individuals as a pathfinding exercise for identifying high-value data on how the human body changes in spaceflight[43].

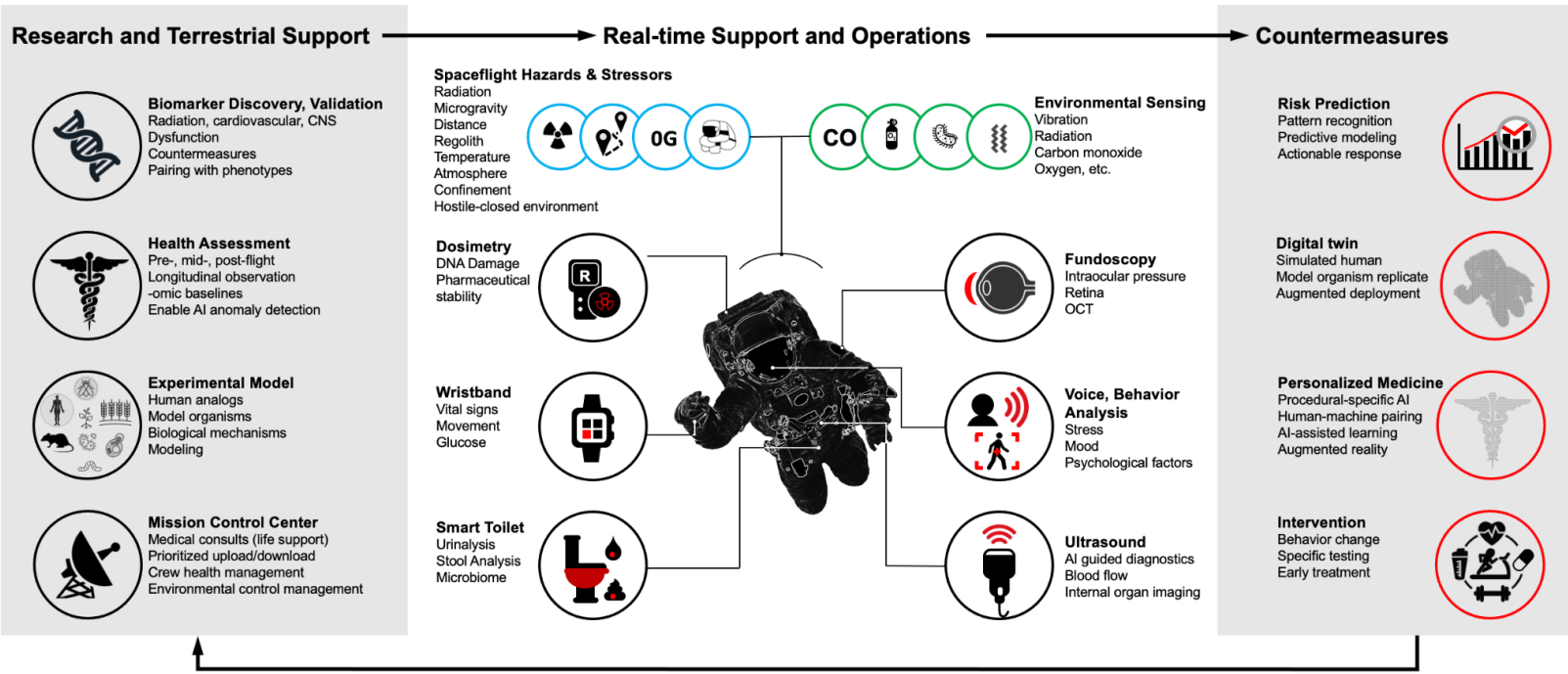

**Figure 1. Precision space health system.** Precision Space Health is an intelligent and research-developed system which: monitors, aggregates, and assess biomedical statuses; analyzes and predicts personalized adverse health outcomes; adapts and responds to newly accumulated data; and provides preventive actionable insights to support the crew medical officer and individual crew.

A healthy and high-performing Astronaut Corps is integral to mission success. Additionally, NASA has an ethical responsibility to minimize (and potentially eliminate) the long-term health and quality of life impacts from spaceflight, a goal for which Precision Space Health is ideally suited. Lastly, technological development for spaceflight provides an opportunity for leadership and substantial technology transfer to terrestrial science and healthcare-business applications, for which NASA already has a long and inspiring history[44].

Spaceflight healthcare needs to evolve toward a Precision Space Health system powered by AI/ML tools for automated assessment and prediction in deep space missions; however, the technologies and techniques for in-mission PM/PH usage have yet to be realized.

Current Approaches with Potential for Precision Space Health

Terrestrial approaches to PM/PH are still maturing and have limited clinical implementation. However, there are developments in these fields where the data collection, analysis, and interpretation of health-related data and biomarkers are making significant headway when coupled with AI/ML approaches[29]. In Table 1, we summarize current technologies to highlight promising paths forward.

| | Category | Technology | Relevance to spaceflight |
|---|---|---|---|
| **Terrestrial** | Routine screening and diagnostics | Deep learning detection of retinopathy and macular degeneration[45,46] | Ocular health risk in spaceflight[47–49] |
| | | Deep learning prediction of cardiovascular disease risk based on retinal photography or clinical data[50–52] | Cardiac health risk in spaceflight[53] |
| | | Deep learning detection and diagnosis of cardiovascular disease from electrocardiograms[54,55] | Cardiac health risk in spaceflight[53] |
| | | Voice analysis methods for detection of changes in mood and mental health state or relationship outcomes[56–58] | Isolation and mental health[59,60] |
| | | Continuous human excreta monitoring through smart toilets[32] | Genitourinary, gastrointestinal health including microbiome assessment |
| | Individualized subtyping and modeling | Omics-based individualized prediction of drug metabolism and effectiveness[61,62] | Spaceflight-impacted pharmaceutical degradation[62] |
| | | Subtyping patients based on genomic markers such as DNA mutations[63,64] | Radiation-induced cancer risk[10] |
| | | Medical digital twins as computational models of individual patients[65] | Drug efficacy prediction |
| | | Tissue-on-a-chip models[66] | Personalized spaceflight risk effect prediction[67,68] |
| | Clinical and decision support | Handheld ultrasound pocket probes with AI-guided diagnostic interpretation[69] | Imaging and evaluation in remote locations |
| | | AI-aided surgical procedures[70] | Surgery in remote locations[71] |
| | | Augment and enhance ultrasound performance[72,73] | Clinical workflow support |
| **Spaceflight or Analog** | Monitoring | Fundoscopy and optical coherence tomography for neuro-ophthalmic physiology[47] | Spaceflight associated neuro-ocular syndrome[74] |
| | | Continuous electrocardiogram recordings during stressful mission events[53] | Cardiac health risk in spaceflight[53] |
| | | Operational analysis tool for estimating organ radiation dosimetric quantities at specific vehicle locations[75] | Radiation-induced disease risk[10] |
| | | Continuous physiological monitoring through skin touch[76] | Collect baseline data and detect anomalies[76] |

**Table 1.** Current state of terrestrial and spaceflight or analog demonstrated approaches to PM/PH or clinical AI/ML implementation with promise for adaptation to a Precision Space Health system.

As evidenced in Table 1, current spaceflight-ready technologies are limited to health monitoring. Available terrestrial technologies are in varying stages of maturity and require additional investment in development on Earth, but show promise and relevance to detect and mitigate spaceflight risks. Each of these fields are advancing individually, some rapidly, some very slowly, and often only indirectly related to the challenges of human spaceflight. These technologies must be developed in parallel and woven together with a common goal in mind if they are ever to support space exploration. We discuss each of these key technologies in detail in the following sections.

**Biomarkers and Health Status Assessment**

Biomarkers

The proposed Precision Space Health system will depend on robust sets of biomarkers for early disease detection and prediction[77–81], shifting the model of crew health from treatment to prevention[82]. This necessitates investment in future terrestrial and space research to identify key biomarkers to predict specific disorders. Biomarkers can be used as markers of dysfunction and also of successful countermeasures. These measured biomarkers must also be safe, reliable, and reproducible in order to assess biomedical function and assess environmental impacts on health[15], and biomarkers must be distinguished between those which are causal versus indicative of a disease[83]. While the suite of clinically actionable biomarkers are still being discovered and implemented, biomarker screening has the potential for improved health maintenance[84,85] through enabling early identification of subclinical changes, which in turn enables early intervention and prevention. As real-time audio/video evaluation capabilities decrease in deep space travel, we propose that biomarker monitoring will crucially make up for these deficiencies by providing a detailed knowledge base regarding the real-time status of human health and biological systems. Further, the development of AI-driven models capable of individualized interpretation of biomarkers will allow further acceptance and establishment of monitoring thresholds on a personalized basis[86]. We discuss several examples of potential spaceflight health biomarkers below.

*Predictions Using Multi-Omic, Paired, and Phenotypic Data*

Considering the known altered immune function effects from spaceflight, the immune system is a key biological system to monitor and evaluate health status. Single-cell immune monitoring approaches such as single-cell RNA sequencing (scRNAseq)[87,88] and mass cytometry (CyTOF)[89] are currently used to monitor populations in terrestrial clinical settings where the immune system is modulated[90,91]. These approaches, combined with comprehensive immune measurements[92] such as levels of circulating cytokines, can be harnessed to map baseline immune configurations of astronauts. These data then can be used by AI models to detect changes in immune competence during spaceflight. Changes in the microbiome during spaceflight could also indirectly affect the immune system. Therefore a multi-omic approach pairing pre-flight, single-cell immune data and microbiome data, could be used to build an integrative model to identify changes in microbial communities as predictive biomarkers for detection of immune system decline utilizing the regularly monitored and analyzed spacecraft and individual microbiomes[91]. This integrative approach could be extended to more broadly associate with other relevant biomarkers such as microRNAs[93], exosomes, cell-free DNA[94] and clonal hematopoiesis[95], and DNA damage responses[96]. Biomarkers based on behavioral phenotypes (speech patterns, semantic/sentiment breakdown, facial expressions) can also be paired with multi-omics data to create powerful (and required) omic-phenotypic connections[97]. This approach can fill a need that was identified on the International

Space Station (ISS) where available countermeasures were shown to improve observed immune dysregulation[98].

*Predictions Using Longitudinal, Individualized, and Baseline Data*

For biomarker research as well as in future real-time space health monitoring, longitudinal measurements will be essential to detect individualized health changes. As the NASA Twins study showed[43], even comparing an astronaut to their twin on earth is limited: the best baseline for an individual is oneself. For example, continuous monitoring of temperature with wearables has shown that fever thresholds change between individuals, or with age, gender, or ethnicity[99]. Similarly, systems-level analyses have shown variations in immune setpoints[90] and microbiome composition[100] across the population along a continuum. Previous efforts have successfully demonstrated individualized monitoring of changes to self-baselines using blood[38], digital devices[39], and non-invasive saliva sampling[41], enabling personalized coaching of individuals[101] and microbiomes[40]. Such approaches can first establish baselines of various biomarkers for each astronaut individually on Earth. AI methods for time-series analysis, particularly utilizing changepoint or anomaly detection, can be used to identify potentially adverse medical events through monitoring deviations from a healthy baseline using longitudinal data[38–41,101]. When digital twin technology is mature, having a digital twin of each astronaut (rather than a biological twin) would aid predictive power[65]. When comparing metrics in space, careful consideration must be taken to understand if observed changes reflect a healthy response or not. Long-term changes over the course of a mission will provide insight into whether an astronaut is slowly drifting to an unhealthy state or simply adjusting to their new environment. Model organism reference experiments and missions can also be a testing bed for longitudinal, individualized, and predictive spaceflight health monitoring.

Health Status Assessment

AI can also be incorporated into health status assessment systems in a non-invasive manner to generate more longitudinal data. For example, AI voice analysis can be used to monitor stress or fatigue, with privacy considerations mitigated by avoiding semantic analysis of dialogue[56]. AI can analyze sleep and locomotion activity, and can assess how inferred health status is affected by various events. To mitigate stress, AI-generated personalized and private therapy programs can be included in crew health resources; for example, immersive virtual reality-based revitalization or AI-CBT chatbots[102,103]. A further example is the pairing of pre-flight structural 'biomarker' (i.e., anatomical) analysis for monitoring vascular and tissue ocular/vision changes [45,47] with assessments of adverse headward fluid shifts occurring in microgravity. Such spaceflight-based results are potentially linked together with other physiological monitoring of blood, nutritional, immunological and performance measures, where further advances would include AI analysis and equipment miniaturization[29,46,50]. Spaceflight assessments can also motivate the importance of acquiring more space data, rather than relying on terrestrial analogs that may not replicate key features of human responses to space[74].

Of significant importance, novel phenotypic manifestations and situations should be expected to occur in deep space, due to the concurrent synergistic interactions between the several known spaceflight health hazards[1,5]. The use of reinforcement learning and n-of-1 studies[79,80,104,105] may help provide statistical power to derive multi-targeted treatment, behavioral interventions or activity interventions (specific to individuals) during a mission in order to address novel phenotypes.

**The Spectrum of Flight Data Acquisition: Layered and Integrated Monitoring**

We propose a multi-layered monitoring approach including both the spacecraft environment and individual astronauts (**Figure 2**) as part of a Precision Space Health system. Both non-contact and contact devices will be used to monitor individual astronauts. Novel semi-automated assays to monitor the entire spacecraft and all habitable environments should be developed. This holistic monitoring would provide for the first time a continuous picture of the health of the entire spacecraft or habitat, and the living ecosystem inside.

The initial layer of monitoring would involve the continuous environmental sensing of physical (vibrations, humidity, temperature, airflow, sounds, electromagnetic radiation, etc.), chemical (carbon dioxide, oxygen, dust particles, volatile organic and inorganic compounds, etc.), and biological (general microbiota and specific species with known health risks) components. Many of these sensors are already standard instruments for LEO missions, so it would be straightforward to begin testing integration of sensor data, ML models, and human in-the-loop inputs[22]. For example, radiation instruments currently in use on the ISS and planned for lunar missions gather real-time data on absorbed radiation dose and dose rates[106–108]. However, extrapolation from absorbed dose to specific biological effects require a detailed knowledge of the components of the space radiation field, which in turn requires post-processing on the ground[106,109]. AI models deployed *in situ* for deep space missions could integrate data and provide real-time estimates of biological radiation risk to assist in the prescription of appropriate physical or biological or pharmaceutical countermeasures, with data also being used for terrestrial long-term health management post-mission. Next-generation sensors with integrated data processing and analytics should be introduced to streamline and enable immediately decipherable metrics. The AI approach of active learning[110] should be considered especially for this first layer of monitoring, which uses intermittent human input and annotation to adapt to changes in the environment and facilitate assessment (uncertainty, diversity, randomness), of constant and large amounts of monitoring data from interdependent environments. These results can be presented to the crew with easily interpretable readouts. These data would also be transferred to the spacecraft or habitat Precision Space Health system for analysis and integration with existing knowledge of the biological effects of environmental stressors.

A second layer is traditional non-invasive physiological metrics, collected by 'wearables', point-of-care devices (e.g., ultrasound, blood pressure, breath-analysis, ocular/visual, respiration), videos indicating behavioral health, and self-administered tests, such as cognitive tests, exercise routines and sleep data[15]. Platforms should be minimally intrusive, data should be easily decipherable to crew and CMO, and data collection should not overly consume crew time. An example of non-intrusive monitoring was recently demonstrated using active sonar (speaker-microphone) to remotely monitor heart rate and heart rhythms[111].

A third layer would be based on molecular-physiological biomarkers and/or truly 'invasive' measures obtained from various swabs, blood draws, saliva sampling and other molecular assays. A 'smart toilet' (as well as smart showerbooth, smart mirror) could preserve and prepare waste specimens for biochemistry assays and microbiome profiling[32,112]. Such platforms hold promise in expanding to include *in situ* and real-time analytical capabilities. Similarly, non-invasive high-frequency monitoring of molecular components from saliva over time can also provide immune signatures that may be used to monitor deviations from a healthy immune baseline, utilizing anomaly detection algorithms to assess changepoints as potential adverse medical events[41]. Paired with an AI-assisted biological knowledge base including expected baselines and biomarkers, such non-invasive approaches could assist in predicting adverse health outcomes and identifying preventive actions. Also of crucial importance for missions beyond the Earth's Van Allen Belt (which conveys a degree of radioprotection), molecular and sophisticated dosimetry will be essential for high resolution detection of both DNA damage[96] and gauging pharmaceutical stability[113,114]. This entire multilayered and integrated monitoring approach would be attractive for crews, as it is less invasive, less cumbersome and encourages more participation with near-immediate and seamless feedback. Such a layered system would increase the accurate monitoring of the true health of the ecosystem and the crew, by developing a more holistic model that integrates multidimensional and multimodal measures.

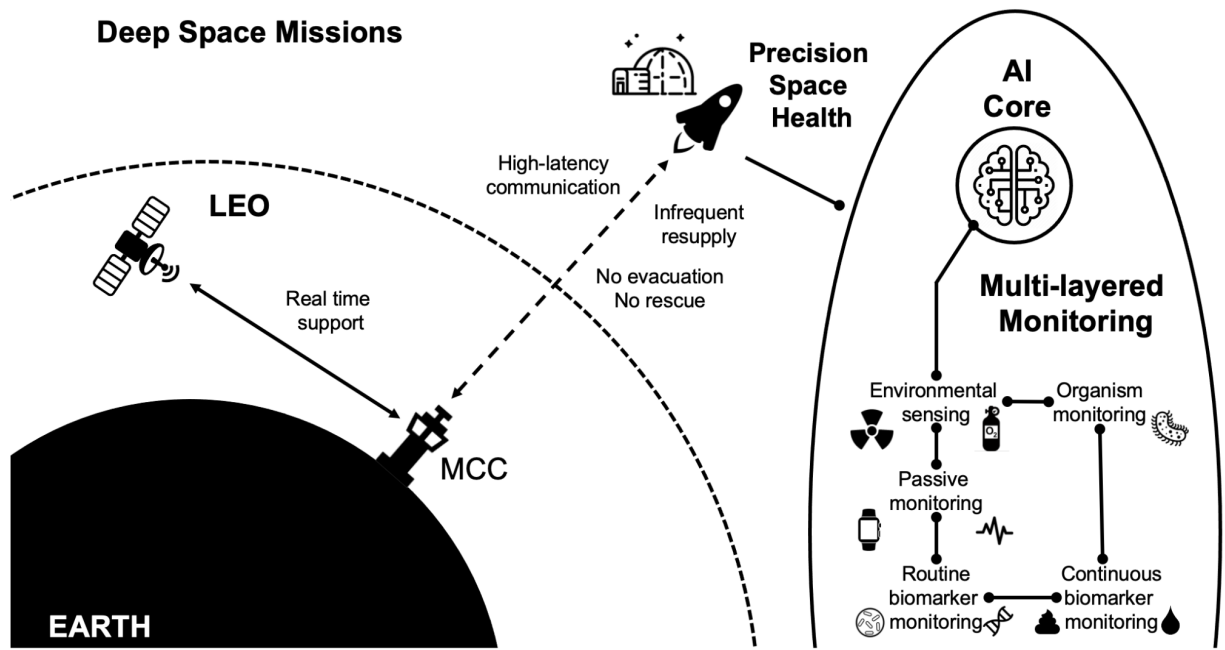

**Figure 2. Layered and integrated data acquisition and monitoring.** A schematic of the layered and integrated biological and health monitoring system, enabling precision astronaut health support during deep space missions with limited Earth communications ability, through a core system of AI-powered monitoring, sensing and prediction.

**Adaptations in Computing, Model Training, and Data Communications in Deep Space**

Historically, computing hardware and data transmission systems to support astronaut health have been predominantly designed for LEO operations, relying on terrestrial computing using downlinked biosensor telemetry catered to a ground-based flight surgeon. As humanity embarks on long-duration cis-Lunar and planetary-class exploration missions[115], the computing hardware and data transmission requirements for astronaut health systems will change, as will the feasibility and approaches for model training in Earth-linked vs *in situ* settings. For Lunar spaceflight and habitats, the volume of health-related data may grow to the point that it is no longer viable to downlink the datasets and they must instead be processed *in situ* using Lunar information technology centers with sufficient computational capacity. For Mars missions, high-latency communications prohibit real-time health support and AI applications must offer near real-time capabilities, operate *in situ* and be maximally autonomous.

Data Communications and Computing

Therefore, the hardware and data communication strategies for astronaut health beyond LEO must adapt to optimally balance the continued efficacy of terrestrial computing for certain needs (e.g., routine monthly blood-work analysis) versus the importance of autonomous AI support for time-sensitive requirements (e.g., ECG monitoring of an astronaut during a Mars surface excursion). For these *in situ* needs, environmental and flight requirements will also impact the design of computing, storage and communication systems[116]. These factors include volume, weight and power constraints, resilience to launch vibration, chemically reactive planetary dusts, ionizing radiation and reliability in the context of autonomous maintenance and repair. The pioneering development of the Spaceborne Computer-1 and -2[117] aboard the ISS has made inroads toward this, along with testing on-board graphics processing units (GPU) for AI and ML capabilities, including real-time base calling that can detect modified nucleic acids[118]. Future AI and data systems need to be designed to minimize the need of transferring data back to Earth, or take advantage of ML-based data compression[119] or active learning systems based on pre-trained, constantly learning ML systems[110], all while working within the limited computing environment of a spacecraft or distant habitat.

Modeling with Deep Space Biomedical Data

With respect to AI/ML models, most current terrestrial AI techniques use large numbers of observations (and usually a large feature set) to train a model. The application of these methods to human health in the context of spaceflight exploration is challenging for several reasons. First, biomedical data collected from astronauts in-flight are historically and currently limited. Just over 600 people worldwide have gone to space, and only 24 beyond LEO. Missions have had an average duration of fewer than 30 days, making it almost impossible to train sophisticated AI/ML models using only spaceflight human data. Second, the data collected during spaceflight have been narrow and inconsistent compared to what is typically available in clinical or research settings on Earth. Until recently, there was no standardized set of biomedical measures taken during NASA human spaceflight explicitly for research purposes[120]. Finally, NASA's mission of exploration means that the models needed for Precision Space Health must extrapolate beyond the context in which we have spaceflight experience. Even though the laws of physics do not change over the course of a deep space mission, the human body does, and potentially in non-linear ways, thus lowering the accuracy of the "approximation" training data. A key consideration for AI medical applications is whether the system needs to be trained *in situ* using locally collected data, or if the model can be trained using ground data prior to the mission, during the mission with an uplink of the updated model, or gradually developed into active learning systems[110] to add another degree of autonomy. These distinctions are important since training a model is typically intensive computationally, and requires large amounts of data, whereas performing inference with a trained model is far less demanding.

Table 2 summarizes the workshop recommendations and considerations for developers, scientists, stakeholders and others dedicated to the realization of AI-modeling systems for space biological research and Precision Space Health.

|  | **Requirement** | **Recommendation** | **Application** |
|---|---|---|---|
| **Computing** | *In-situ* data analysis | - Edge computing[13]<br>- Active learning[110] | - Process and analyze data collected in deep space missions on board for input to the Precision Space Health system.<br>- Train and deploy a model, which continuously monitors and retrains itself with self-assessments and regular human inspection. |
|  | Training on distant data | Federated learning[121] | Train a model on data collected in a deep space mission and on Earth-based data for stronger inference. |
|  | Heavy computing needs | - Transfer learning<br>- Dimensionality reduction[121]<br>- tinyML[122]<br>- Few-shot learning[123] | - Train large models on Earth and deploy on data collected in-flight<br>- Identify key features to reduce data size<br>- Prune large neural networks to deploy on spacecraft or habitats with operational constraints<br>- Learn from few data points by leveraging contextual information |
| **Hardware** | Environmental factors (radiation, acceleration, vibration) | Neuromorphic processors | Space-borne computing with very low power, little or no cooling, high efficacy for AI algorithms and resilience to radiation[124,125] |
|  | Monitoring network | Integration with core flight systems | Spectrum of layered biomedical space data acquisition through interconnected personal, nutrition, health objects interconnected with spacecraft network (e.g., ECLSS), with data sharing into the Precision Space Health system (similar to Internet of Things[126]) |
| **Models** | Methods to train on data that differs from inferencing context | Translation[127,128] | e.g., Train on radiation exposure data in animals and predict radiation risks for human crew members. |
|  | Methods for when inferencing data are extremely different (e.g. outliers) to training data | Generalization:<br>- Risk Extrapolation[129,130]<br>- Domain Invariant Representation Learning[129,130] | Prediction in a situation where an astronaut's biosensor data are outliers compared to the terrestrial clinical data used for model training. |
|  | Methods for when inferencing data are persistently different from training data | Adaptation | e.g., Adapting a model trained using terrestrial electrocardiogram data to "new normal" of electrocardiogram readings from astronauts whose heart physiology has changed in spaceflight. |

**Table 2.** Computing, hardware and model development requirements and recommendations for AI-modeling for health in deep space.

The computing hardware, model software, and data management and communication strategies for Precision Space Health need to be maximally adaptive to accommodate constant reassessment from newly acquired data. This is challenged by the need for high data security, as well as spacecraft mission constraints on mass, power, volume and data bandwidth. The feasibility of such a system relies on several factors. First, the predictive maturity of relevant biomarkers and data must be identified and prioritized for sensor development. Second, clinical (personalized) thresholds that can alert when an individual astronaut may be approaching a preventable health issue threshold must be identified and validated. Third, the implementation of threshold-based health assessment must be operationalized for *in situ* analysis in the context of the astronauts involved in a particular spaceflight mission and time course.

Overall, this system will be supported by proactive prioritization of data return. Current communication bandwidth estimates for medical needs do not consider AI analytics and will need to be updated. Research should determine which data type is deemed absolutely essential and mission-critical for Earth-return, to enable flight surgeons and scientists at MCC to provide support (and for researchers to analyze). Further development in ML data compression techniques will aid this process. For interpretable real-time data and analytics, as well as *post hoc* data sharing to Earth (i.e., likely transferring data via laser communications relay or radio frequency[131]), the development of these biomedical monitoring, data acquisition and knowledge extraction systems will rely on pre-defined, robust metadata structures, ontologies, and transfer to data sharing-analytical repositories on Earth. This is covered in the separate companion biological research review article reporting on the workshop (Sanders et al., 2021 [unpublished preprint]).

**Discussion**

Integrated biomedical flight data acquisition, AI-modeling tools and techniques, as well as a Precision Space Health system will be crucial pillars in bridging the gap between our current LEO operational paradigm and that which is needed for successful cis-Lunar and planetary-class missions. The crew's need for progressive independence from Earth, in terms of health and biological self-sufficiency, is largely an informational problem. Workshop participants agreed that all avenues of data, technology, and technique development ought to be explored (not solely AI/ML).

Forward-Looking Questions

How much degradation of crew health and capability should we expect? How much long-term health risk will astronauts face post-career? These questions can only be answered by scientific research, and that is reliant on strong data systems in future missions which can collect, analyze and enable interpretation of results. How much decision support and data analytics capability should be built into a mission to maximize chances of success? This question is limited by current and projected space data capabilities, AI-modeling capabilities and a dearth of opportunities for human systems integration and testing.

Crew Confidence and Fidelity of AI

Data and AI alone does not render clinical care. Data must be interpreted through the lens of clinical significance and used to inform clinical decisions about preventive and acute care. It is important that we shift the role of diagnostic AI from simply predicting labels, to interpreting context and providing iterative cues that guide the diagnostician[132]. Training and building confidence in an AI-based Precision Space Health system must be wholeheartedly established with the crew, flight surgeons and all related staff. Tools and techniques for AI and modeling (or any type of data system) additionally ought to include comprehensive and potentially continuous assessment of its credibility, ethics, and trustworthiness. This includes methods that address reference data or model prediction benchmarks. Fidelity and ethics assessments of AI-modeling extend past specific technique validation. It encompasses broader aspects of credibility including a full provenance of its life cycle development and evaluation of its systemic assumptions, biases, and deployment limitations (i.e., toward predictions or knowledge-gained)[133,134].

Ethics, Genomics, and De-identification

The science, spaceflight and medical communities have a responsibility to meet ethical obligations involved in a Precision Space health system and related data privacy[35,135]. Whole-genome sequencing will likely be a pre-flight component required to enable an effective AI-driven Precision Space Health system. Clear governance, policy and sincere care must be taken to handle the privacy and wishes of all spacefarers (NASA Astronauts, but also international and private-commercial space travelers). Deidentified data systems with decoupled federated learning systemic firewalls are one approach to ensure data are explicitly not traceable[136]. For broad data analysis to occur to support deep space health, completely untraceable data must be attained to protect any impacts on multi-generational offspring to their privacy and quality of life. Development of modified Genetic Information Nondiscrimination Act (GINA) guidelines[35–42,137] and waivers for utilization of spaceflight genomics data, should be considered for the rights of astronauts and their relatives, with the understanding that new guidelines may need to be developed in the future as technology progresses and is incorporated into space missions.

Translational Science and the AI Biomedical Lifecycle

Knowledge and data from both fundamental and applied space biomedical research is part of a crucial translational pipeline to inform a Precision Space Health system. Such research builds a wealth of evidence and statistical power upon which AI biomedical predictions rely. The lifecycle of AI/ML and the cross-cutting relationship between space biological research and Precision Space Health is presented in **Figure 3**.

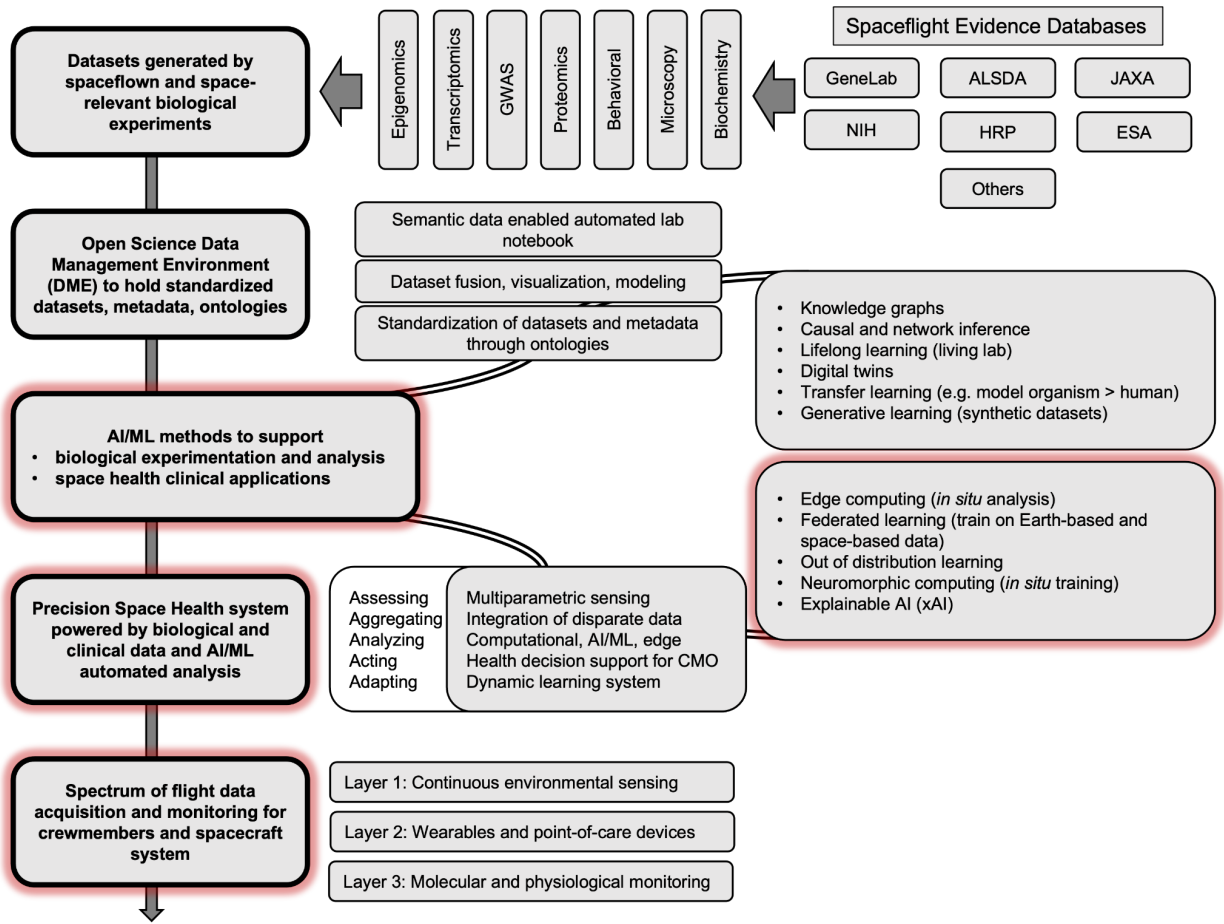

**Figure 3. Space biology and precision space health AI/ML life cycle.** The lifecycle of AI/ML in space biology research and spaceflight health enables full utilization of Open and FAIR space biological datasets through a Data Management Environment. AI/ML methods such as neuromorphic and edge computing, out of distribution learning and explainable AI facilitate a layered and integrated spectrum of in-flight data acquisition to power a Precision Space Health system for deep space missions.

Make Data AI-Ready

Metadata and data curation-processing standards for Precision Space Health and the field of space biology need to be determined, ensuring data are 'AI-ready' and modeling accessible. A space Data Readiness Level metric can be developed as a tool to encourage reliable data quality[138]. Basic synthetic datasets and model libraries for space health and biology also need development, and are part of enabling broad participation by computer scientists, biologists, and algorithmic developers. It is worth considering that AI and modeling approaches of established 'big data' companies and academics may not be suitable for space challenges. Adaptations and innovations in statistics, algorithms, data, and medical informatics are almost certainly going to be required and modified for spaceflight health and biology[139] (Sanders et al., 2021[unpublished preprint]).

Make AI Space-Ready

It will also be important to develop AI approaches that are spaceflight-ready. The unique communication and data transfer challenges of spaceflight are unlike those encountered on Earth for computing and AI-based paradigms. Communication at some of the most important times for health may be disrupted, such as in the context of a solar storm. As a mission moves further from Earth, access to computing power will need to increasingly transition from terrestrial to spacecraft and habitat-based. Many challenges of building space-ready AI are difficult to foresee, and there will be more challenges for Mars missions[140].

Interdisciplinary Teams, Collaboration, and Cross-Cutting Between Research, Engineering, and Clinical

Deep space missions will have one-of-a-kind space biology and health requirements. The interdisciplinary breadth of the teams required to develop, collaborate, refine, and implement these systems is novel. For example, bioinformaticians, clinicians, biologists, algorithmists, data system architects, medical informaticians, engineers, data curators, computer scientists, programmers, are only a few of the categories to consider. The workshop was organized with this interdisciplinary framework in mind, with participants spanning four general domains as seen in **Figure 4**.

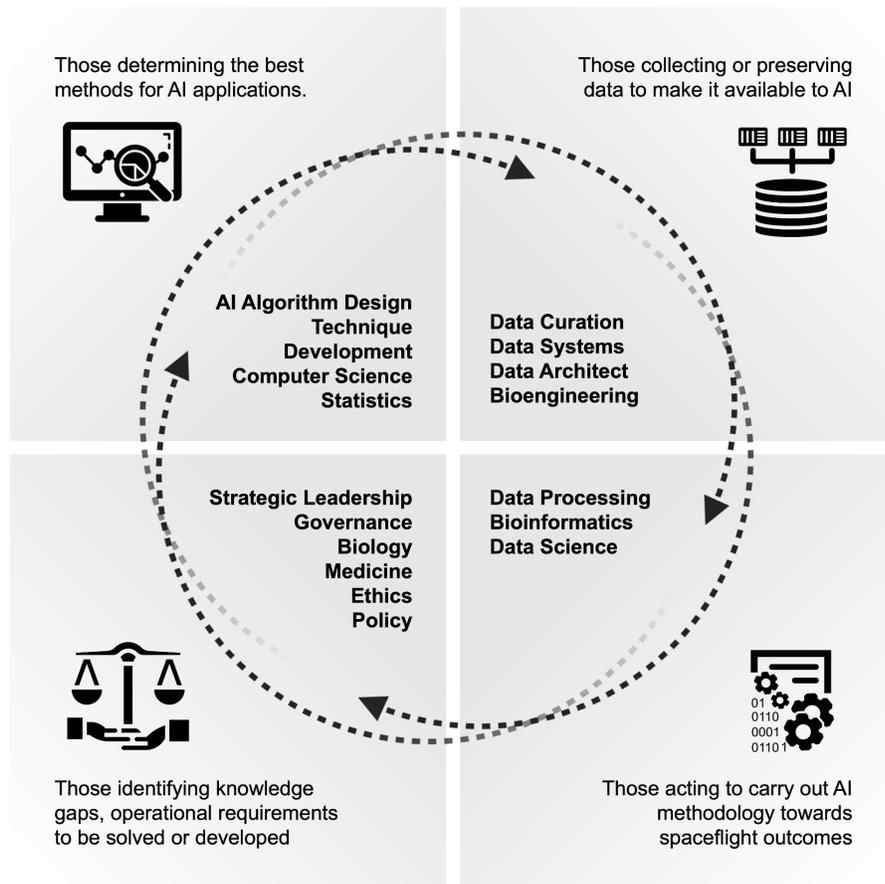

**Figure 4. Novel interdisciplinary collaboration and teams.** The June 2021 "Workshop on Artificial Intelligence and Modeling for Space Biology" was organized to bring together expertise across disciplines. It demonstrates the uniquely required composition and collaboration of teams necessary to develop systems toward space biology and health.

Additionally, to effectively implement all components of Precision Space Health and biological research systems, developers must be collaboratively interwoven so biological researchers, data architects, operational clinicians, and hardware engineers are not siloed within their own domains. Compared to terrestrial and LEO settings, information will likely flow more freely between research and clinical operations during deep space missions, which could result in novel discovery outcomes. For example, a recent research team utilizing an ultrasound device was interested in blood flow behavior in the internal jugular vein in long-term ISS mission crew. The team noticed anomalies in venous blood flow, which in turn led to a clinical team hand-over who proceeded to fully characterize (and treat for) flow stasis and venous thrombosis[141,142]. The integrated data architectures for CHP and the Precision Space Health system are an enabling technology with two goals: (1) improved scientific understanding of the mechanisms of deterioration, and (2) targeted monitoring, prevention and interventionist health countermeasures to ensure mission success[6,143]. In other words, the same data systems that collect and analyze health and performance related information for research purposes can also provide monitoring and decision support to feed back the best information for operational needs.

**Recommendations and Conclusion**

Workshop participants agreed on key technologies for development to enhance deep space biological research and to support space health:
- Biomonitoring Technology
- Biomarker Science
- Mission Implemented Hardware
- Informatic-Algorithmic Software
- Precision Space Health system

The development of the information, data, and AI-modeling systems discussed in this paper will be multi-year, interdisciplinary, and involve far-reaching collaborations across science, engineering, medicine, and operations. As humanity explores beyond LEO, it will leave the confines of an immediately accessible, large, and continuously supportive cohort of mission control health and science staff, with systems that have been developed over the past ~60 years. As deep space missions must be light, robust, agile, and maximally autonomous, the relatively recent development of terrestrial AI, ML, and modeling tools offer a key contribution toward making humanity multi-planetary through spacecraft and habitat biomedical science support and a Precision Space Health system.

**Acknowledgments**

We thank all June 2021 participants and speakers at the "NASA Workshop on Artificial Intelligence & Modeling for Space Biology." Thanks to the NASA Space Biology Program, part of the NASA Biological and Physical Sciences Division within the NASA Science Mission Directorate; as well as the NASA Human Research Program (HRP). Also, thanks to the Space Biosciences Division and Space Biology at Ames Research Center (ARC), especially Diana Ly, Rob Vik, and Parag Vaishampayan. Thanks also to the support provided by NASA GeneLab and the NASA Ames Life Sciences Data Archive. Additional thanks to Sharmila Bhattacharya, NASA Space Biology Program Scientist; Kara Martin, ARC Lead of Exploration Medical Capability (an Element of HRP); as well as Laura Lewis, ARC NASA HRP Lead.

Funding: S.V.C. is funded by NASA Human Research Program grants NNJ16HP24I. S.E.B holds the Heidrich Family and Friends endowed Chair in Neurology at UCSF. S.E.B. also holds the Distinguished Professorship I in Neurology at UCSF. S.E.B is funded by an NSF Convergence Accelerator award (2033569) and NIH/NCATS Translator award (1OT2TR003450). G.I.M was supported by the Translational Research Institute for Space Health, through NASA NNX16AO69A (Project Number T0412). E.L.A. was supported by the Translational Research Institute for Space Health, through NASA NNX16AO69A. C.E.M. thanks NASA grants NNX14AH50G and NNX17AB26G. J.Y. is funded by NIH grant # R00 GM118907. This work was also part of the DOE Agile BioFoundry (http://agilebiofoundry.org), supported by the U.S. Department of Energy, Energy Efficiency and Renewable Energy, Bioenergy Technologies Office, and the DOE Joint BioEnergy Institute (http://www.jbei.org), supported by the Office of Science, Office of Biological and Environmental Research, through contract DE-AC02- 05CH11231 between Lawrence Berkeley National Laboratory and the U.S. Department of Energy. SVK is funded by the Canadian Space Agency (19HLSRM04) and Natural Sciences and Engineering Research Council (NSERC, RGPIN-288253).

**Authorship Contributions**

All authors contributed ideas and discussion during the joint workshop writing session or were speakers at the: "NASA Workshop on Artificial Intelligence & Modeling for Space Biology." R.T.S., L.M.S., and S.V.C. prepared the manuscript. All authors provided feedback on the manuscript.

# References

1. Afshinnekoo, E. *et al.* Fundamental Biological Features of Spaceflight: Advancing the Field to Enable Deep-Space Exploration. *Cell* **183**, 1162–1184 (2020).
2. Loftus, D. J., Rask, J. C., McCrossin, C. G. & Tranfield, E. M. The Chemical Reactivity of Lunar Dust: From Toxicity to Astrobiology. *Earth Moon Planets* **107**, 95–105 (2010).
3. Paul, A.-L. & Ferl, R. J. The biology of low atmospheric pressure -- Implications for exploration mission design and advanced life support. **19**, (2005).
4. Council, N. R. *Recapturing a Future for Space Exploration: Life and Physical Sciences Research for a New Era*. (The National Academies Press, 2011).
5. Goswami, N. *et al.* Maximizing information from space data resources: a case for expanding integration across research disciplines. *Eur. J. Appl. Physiol.* **113**, 1645–1654 (2013).
6. McGuire, K. *et al.* Using systems engineering to develop an integrated crew health and performance system to mitigate risk for human exploration missions. *50th International Conference on Environmental Systems* (12-15 July 2021).
7. Antonsen, E., Hanson, A., Shah, R., Reed, R. D. & Canga, M. A. Conceptual Drivers for an Exploration Medical System. *67th International Astronautical Congress* (2016).
8. Ball, J. R. & Evans, C. H., Jr. *Safe Passage: Astronaut Care for Exploration Missions*. (National Academies Press (US), 2014).
9. McNulty, M. J. *et al.* Evaluating the Cost of Pharmaceutical Purification for a Long-Duration Space Exploration Medical Foundry. *Front. Microbiol.* **12**, 700863 (2021).
10. Blue, R. S. *et al.* Challenges in Clinical Management of Radiation-Induced Illnesses During Exploration Spaceflight. *Aerosp Med Hum Perform* **90**, 966–977 (2019).
11. Chancellor, J. C. *et al.* Limitations in predicting the space radiation health risk for exploration astronauts. *NPJ Microgravity* **4**, 8 (2018).
12. Patel, Z. S. *et al.* Red risks for a journey to the red planet: The highest priority human health risks for a mission to Mars. *NPJ Microgravity* **6**, 33 (2020).
13. McGregor, C. A platform for real-time online health analytics during spaceflight. in *2013 IEEE Aerospace Conference* 1–8 (2013).
14. Hylton, A., Raible, D. & Clark, G. A Delay Tolerant Networking-Based Approach to a High Data Rate Architecture for Spacecraft. in *2019 IEEE Aerospace Conference* 1–10 (2019).
15. Strangman, G. E. *et al.* Deep-space applications for point-of-care technologies. *Current Opinion in Biomedical Engineering* **11**, 45–50 (2019).
16. Walton, M. E. & Kerstman, E. L. Quantification of Medical Risk on the International Space Station Using the Integrated Medical Model. *Aerosp Med Hum Perform* **91**, 332–342 (2020).
17. Sipes, W., Holland, A. & Beven, G. Managing Behavioral Health in Space. in *Handbook of Bioastronautics* (eds. Young, L. R. & Sutton, J. P.) 425–436 (Springer International Publishing, 2021).
18. Thompson, D. E. *Space Technology - Game Changing Development NASA Facts: Autonomous Medical Operations*. https://ntrs.nasa.gov/citations/20180002541 (2018).
19. Castro-Wallace, S. L. *et al.* Nanopore DNA Sequencing and Genome Assembly on the International Space Station. *Sci. Rep.* **7**, 18022 (2017).
20. McGregor, C. A Platform for Real-Time Space Health Analytics as a Service Utilizing Space Data Relays. in *2021 IEEE Aerospace Conference (50100)* 1–14 (2021).
21. Mindock, J. *et al.* Systems Engineering for Space Exploration Medical Capabilities. in *AIAA SPACE and Astronautics Forum and Exposition* (American Institute of Aeronautics and Astronautics, 2017).
22. Schneider, W. F. & Perry, J. L. NASA environmental control and life support technology development and maturation for exploration: 2019 to 2020 overview.

https://ttu-ir.tdl.org/bitstream/handle/2346/86395/ICES-2020-200.pdf.
23. Broyan, J. L., Shaw, L., Mc Kinley, M., Meyer, C. & Ewert, M. K. NASA environmental control and life support technology development for exploration: 2020 to 2021 overview. https://ntrs.nasa.gov/api/citations/20210010866/downloads/ICES_384-FY2021%20ECLSS%20Overview-1676Review%20-%20Final.docx.pdf.
24. Williams-Byrd, J. A. *et al.* Implementing NASA?s Capability-Driven Approach: Insight into NASA?s Processes for Maturing Exploration Systems. in *AIAA SPACE 2015 Conference and Exposition* (American Institute of Aeronautics and Astronautics, 2015).
25. Jordan, M. I. & Mitchell, T. M. Machine learning: Trends, perspectives, and prospects. *Science* **349**, 255–260 (2015).
26. Hood, L. & Flores, M. A personal view on systems medicine and the emergence of proactive P4 medicine: predictive, preventive, personalized and participatory. *N. Biotechnol.* **29**, 613–624 (2012).
27. Zitnik, M. *et al.* Machine Learning for Integrating Data in Biology and Medicine: Principles, Practice, and Opportunities. *Inf. Fusion* **50**, 71–91 (2019).
28. Yu, K.-H., Beam, A. L. & Kohane, I. S. Artificial intelligence in healthcare. *Nat Biomed Eng* **2**, 719–731 (2018).
29. Topol, E. *Deep Medicine: How Artificial Intelligence Can Make Healthcare Human Again*. (Basic Books).
30. Topol, E. J. High-performance medicine: the convergence of human and artificial intelligence. *Nat. Med.* **25**, 44–56 (2019).
31. National Research Council (US) Committee on A Framework for Developing a New Taxonomy of Disease. *Toward Precision Medicine: Building a Knowledge Network for Biomedical Research and a New Taxonomy of Disease*. (National Academies Press (US), 2012).
32. Park, S.-M., Ge, T. J., Won, D. D., Lee, J. K. & Liao, J. C. Digital biomarkers in human excreta. *Nat. Rev. Gastroenterol. Hepatol.* **18**, 521–522 (2021).
33. Schmidt, M. A., Schmidt, C. M., Hubbard, R. M. & Mason, C. E. Why Personalized Medicine Is the Frontier of Medicine and Performance for Humans in Space. *New Space* **8**, 63–76 (2020).
34. Kahn, J., Liverman, C. T. & McCoy, M. A. *Health Standards for Long Duration and Exploration Spaceflight: Ethics Principles, Responsibilities, and Decision Framework*. (The National Academies Press, 2014).
35. Antonsen, E. L. & Reed, R. D. Policy considerations for precision medicine in human spaceflight. https://www.law.uh.edu/hjhlp/volumes/Vol_19/1%20-%20Antonsen%20and%20Reed%20(pp%201-37).pdf.
36. Schork, N. J. Personalized medicine: Time for one-person trials. *Nature* **520**, 609–611 (2015).
37. Arges, K. *et al.* The Project Baseline Health Study: a step towards a broader mission to map human health. *NPJ Digit Med* **3**, 84 (2020).
38. Chen, R. *et al.* Personal omics profiling reveals dynamic molecular and medical phenotypes. *Cell* **148**, 1293–1307 (2012).
39. Li, X. *et al.* Digital Health: Tracking Physiomes and Activity Using Wearable Biosensors Reveals Useful Health-Related Information. *PLoS Biol.* **15**, e2001402 (2017).
40. Zhou, W. *et al.* Longitudinal multi-omics of host-microbe dynamics in prediabetes. *Nature* **569**, 663–671 (2019).
41. Mias, G. I. *et al.* Longitudinal saliva omics responses to immune perturbation: a case study. *Sci. Rep.* **11**, 710 (2021).
42. Haney, N. M., Urman, A., Waseem, T., Cagle, Y. & Morey, J. M. AI's role in deep space. *J. Med. Artif. Intell.* **3**, 11–11 (2020).
43. Garrett-Bakelman, F. E. *et al.* The NASA Twins Study: A multidimensional analysis of a year-long human spaceflight. *Science* **364**, (2019).

44. NASA Technology Transfer Program. Spinoff 2019.
45. Gulshan, V. *et al.* Development and Validation of a Deep Learning Algorithm for Detection of Diabetic Retinopathy in Retinal Fundus Photographs. *JAMA* **316**, 2402–2410 (2016).
46. De Fauw, J. *et al.* Clinically applicable deep learning for diagnosis and referral in retinal disease. *Nat. Med.* **24**, 1342–1350 (2018).
47. Lee, A. G. *et al.* Spaceflight associated neuro-ocular syndrome (SANS) and the neuro-ophthalmologic effects of microgravity: a review and an update. *NPJ Microgravity* **6**, 7 (2020).
48. Vyas, R. J. *et al.* Decreased Vascular Patterning in the Retinas of Astronaut Crew Members as New Measure of Ocular Damage in Spaceflight-Associated Neuro-ocular Syndrome. *Invest. Ophthalmol. Vis. Sci.* **61**, 34 (2020).
49. Lagatuz, M. *et al.* Vascular Patterning as Integrative Readout of Complex Molecular and Physiological Signaling by VESsel GENeration Analysis. *J. Vasc. Res.* **58**, 207–230 (2021).
50. Poplin, R. *et al.* Prediction of cardiovascular risk factors from retinal fundus photographs via deep learning. *Nat Biomed Eng* **2**, 158–164 (2018).
51. Weng, S. F., Reps, J., Kai, J., Garibaldi, J. M. & Qureshi, N. Can machine-learning improve cardiovascular risk prediction using routine clinical data? *PLoS One* **12**, e0174944 (2017).
52. Paschalidis, Y. How Machine Learning Is Helping Us Predict Heart Disease and Diabetes. *Harvard Business Review* (2017).
53. Lee, S. M. C., Stenger, M. B., Laurie, S. S. & Macias, B. R. Evidence Report: Risk of Cardiac Rhythm Problems During Spaceflight. *NASA Human Research Roadmap*.
54. Strodthoff, N. & Strodthoff, C. Detecting and interpreting myocardial infarction using fully convolutional neural networks. *Physiol. Meas.* **40**, 015001 (2019).
55. Hannun, A. Y. *et al.* Cardiologist-level arrhythmia detection and classification in ambulatory electrocardiograms using a deep neural network. *Nat. Med.* **25**, 65–69 (2019).
56. Or, F., Torous, J. & Onnela, J.-P. High potential but limited evidence: Using voice data from smartphones to monitor and diagnose mood disorders. *Psychiatr. Rehabil. J.* **40**, 320–324 (2017).
57. Nasir, M., Baucom, B. R., Georgiou, P. & Narayanan, S. Predicting couple therapy outcomes based on speech acoustic features. *PLoS One* **12**, e0185123 (2017).
58. Frankel, J. How Artificial Intelligence Could Help Diagnose Mental Disorders. *The Atlantic* (2016).
59. Landon, L. B., Slack, K. J. & Barrett, J. D. Teamwork and collaboration in long-duration space missions: Going to extremes. *Am. Psychol.* **73**, 563–575 (2018).
60. Willams, R. S. & Davis, J. R. A critical strategy: ensuring behavioral health during extended-duration space missions. *Aviat. Space Environ. Med.* **76**, B1–2 (2005).
61. Stingl, J. C., Welker, S., Hartmann, G., Damann, V. & Gerzer, R. Where Failure Is Not an Option -Personalized Medicine in Astronauts. *PLoS One* **10**, e0140764 (2015).
62. Blue, R. S. *et al.* Supplying a pharmacy for NASA exploration spaceflight: challenges and current understanding. *NPJ Microgravity* **5**, 14 (2019).
63. Ashley, E. A. Towards precision medicine. *Nat. Rev. Genet.* **17**, 507–522 (2016).
64. Wesseling, P. & Capper, D. WHO 2016 Classification of gliomas. *Neuropathol. Appl. Neurobiol.* **44**, 139–150 (2018).
65. Masison, J. *et al.* A modular computational framework for medical digital twins. *Proc. Natl. Acad. Sci. U. S. A.* **118**, (2021).
66. Low, L. A., Mummery, C., Berridge, B. R., Austin, C. P. & Tagle, D. A. Organs-on-chips: into the next decade. *Nat. Rev. Drug Discov.* **20**, 345–361 (2021).
67. Tissue Chips in Space. https://ncats.nih.gov/tissuechip/projects/space (2016).
68. Yeung, C. K. *et al.* Tissue Chips in Space-Challenges and Opportunities. *Clin. Transl. Sci.* **13**, 8–10 (2020).

69. Baribeau, Y. *et al.* Handheld Point-of-Care Ultrasound Probes: The New Generation of POCUS. *J. Cardiothorac. Vasc. Anesth.* **34**, 3139–3145 (2020).
70. Hashimoto, D. A., Rosman, G., Rus, D. & Meireles, O. R. Artificial Intelligence in Surgery: Promises and Perils. *Ann. Surg.* **268**, 70–76 (2018).
71. Haidegger, T., Sándor, J. & Benyó, Z. Surgery in space: the future of robotic telesurgery. *Surg. Endosc.* **25**, 681–690 (2011).
72. Akkus, Z. *et al.* A Survey of Deep-Learning Applications in Ultrasound: Artificial Intelligence-Powered Ultrasound for Improving Clinical Workflow. *J. Am. Coll. Radiol.* **16**, 1318–1328 (2019).
73. Bowness, J., Varsou, O., Turbitt, L. & Burkett-St Laurent, D. Identifying anatomical structures on ultrasound: assistive artificial intelligence in ultrasound-guided regional anesthesia. *Clin. Anat.* **34**, 802–809 (2021).
74. Taibbi, G. *et al.* Opposite response of blood vessels in the retina to 6° head-down tilt and long-duration microgravity. *NPJ Microgravity* **7**, 38 (2021).
75. Mertens, C. J., Slaba, T. C. & Hu, S. Active dosimeter-based estimate of astronaut acute radiation risk for real-time solar energetic particle events. *Space Weather* **16**, 1291–1316 (2018).
76. Toscano, W. *et al. Wearable Biosensor Monitor to Support Autonomous Crew Health and Readiness to Perform*. https://ntrs.nasa.gov/citations/20190001996 (2017).
77. Gambhir, S. S., Ge, T. J., Vermesh, O. & Spitler, R. Toward achieving precision health. *Sci. Transl. Med.* **10**, (2018).
78. Gambhir, S. S., Ge, T. J., Vermesh, O., Spitler, R. & Gold, G. E. Continuous health monitoring: An opportunity for precision health. *Sci. Transl. Med.* **13**, (2021).
79. Goel, N. & Dinges, D. F. Predicting Risk in Space: Genetic Markers for Differential Vulnerability to Sleep Restriction. *Acta Astronaut.* **77**, 207–213 (2012).
80. Limkakeng, A. T., Jr *et al.* Systematic Molecular Phenotyping: A Path Toward Precision Emergency Medicine? *Acad. Emerg. Med.* **23**, 1097–1106 (2016).
81. Clément, G. R. *et al.* Challenges to the central nervous system during human spaceflight missions to Mars. *J. Neurophysiol.* **123**, 2037–2063 (2020).
82. Schüssler-Fiorenza Rose, S. M. *et al.* A longitudinal big data approach for precision health. *Nat. Med.* **25**, 792–804 (2019).
83. Budd, S. *et al.* Prototyping CRISP: A Causal Relation and Inference Search Platform applied to Colorectal Cancer Data. in *2021 IEEE 3rd Global Conference on Life Sciences and Technologies (LifeTech)* 517–521 (2021).
84. Fitzgerald, J. *et al.* Future of biomarker evaluation in the realm of artificial intelligence algorithms: application in improved therapeutic stratification of patients with breast and prostate cancer. *J. Clin. Pathol.* **74**, 429–434 (2021).
85. Weiss, J., Hoffmann, U. & Aerts, H. J. W. L. Artificial intelligence-derived imaging biomarkers to improve population health. *The Lancet. Digital health* vol. 2 e154–e155 (2020).
86. Schmidt, M. A. & Goodwin, T. J. Personalized medicine in human space flight: using Omics based analyses to develop individualized countermeasures that enhance astronaut safety and performance. *Metabolomics* **9**, 1134–1156 (2013).
87. Papalexi, E. & Satija, R. Single-cell RNA sequencing to explore immune cell heterogeneity. *Nat. Rev. Immunol.* **18**, 35–45 (2018).
88. Gertz, M. L. *et al.* Multi-omic, Single-Cell, and Biochemical Profiles of Astronauts Guide Pharmacological Strategies for Returning to Gravity. *Cell Rep.* **33**, 108429 (2020).
89. Spitzer, M. H. & Nolan, G. P. Mass Cytometry: Single Cells, Many Features. *Cell* **165**, 780–791 (2016).
90. Lakshmikanth, T. *et al.* Human Immune System Variation during 1 Year. *Cell Rep.* **32**, 107923

(2020).
91. Hartmann, F. J. *et al.* Comprehensive Immune Monitoring of Clinical Trials to Advance Human Immunotherapy. *Cell Rep.* **28**, 819–831.e4 (2019).
92. Emerson, R. O. *et al.* Immunosequencing identifies signatures of cytomegalovirus exposure history and HLA-mediated effects on the T cell repertoire. *Nat. Genet.* **49**, 659–665 (2017).
93. Malkani, S. *et al.* Circulating miRNA Spaceflight Signature Reveals Targets for Countermeasure Development. *Cell Rep.* **33**, 108448 (2020).
94. Bezdan, D. *et al.* Cell-free DNA (cfDNA) and Exosome Profiling from a Year-Long Human Spaceflight Reveals Circulating Biomarkers. *iScience* **23**, 101844 (2020).
95. Mencia-Trinchant, N. *et al.* Clonal Hematopoiesis Before, During, and After Human Spaceflight. *Cell Rep.* **33**, 108458 (2020).
96. Pariset, E. *et al.* DNA Damage Baseline Predicts Resilience to Space Radiation and Radiotherapy. *Cell Rep.* **33**, 108434 (2020).
97. Bruce-Keller, A. J., Salbaum, J. M. & Berthoud, H.-R. Harnessing Gut Microbes for Mental Health: Getting From Here to There. *Biol. Psychiatry* **83**, 214–223 (2018).
98. Crucian, B. E. *et al.* Countermeasures-based Improvements in Stress, Immune System Dysregulation and Latent Herpesvirus Reactivation onboard the International Space Station - Relevance for Deep Space Missions and Terrestrial Medicine. *Neurosci. Biobehav. Rev.* **115**, 68–76 (2020).
99. Obermeyer, Z., Samra, J. K. & Mullainathan, S. Individual differences in normal body temperature: longitudinal big data analysis of patient records. *BMJ* **359**, j5468 (2017).
100. Manor, O. *et al.* Health and disease markers correlate with gut microbiome composition across thousands of people. *Nat. Commun.* **11**, 5206 (2020).
101. Price, N. D. *et al.* A wellness study of 108 individuals using personal, dense, dynamic data clouds. *Nat. Biotechnol.* **35**, 747–756 (2017).
102. Gratzer, D. & Goldbloom, D. Therapy and E-therapy-Preparing Future Psychiatrists in the Era of Apps and Chatbots. *Acad. Psychiatry* **44**, 231–234 (2020).
103. Gaffney, H., Mansell, W. & Tai, S. Conversational Agents in the Treatment of Mental Health Problems: Mixed-Method Systematic Review. *JMIR Ment Health* **6**, e14166 (2019).
104. Wang, Y. & Schork, N. J. Power and Design Issues in Crossover-Based N-Of-1 Clinical Trials with Fixed Data Collection Periods. *Healthcare (Basel)* **7**, (2019).
105. Schork, N. J. & Goetz, L. H. Single-Subject Studies in Translational Nutrition Research. *Annu. Rev. Nutr.* **37**, 395–422 (2017).
106. Schimmerling, W. Space radiation dosimetry. *NASA JSC* (2009).
107. Yasuda, H. Effective dose measured with a life size human phantom in a low Earth orbit mission. *J. Radiat. Res.* **50**, 89–96 (2009).
108. Kroupa, M. *et al.* A semiconductor radiation imaging pixel detector for space radiation dosimetry. *Life Sci. Space Res.* **6**, 69–78 (2015).
109. Horneck, G. Biological monitoring of radiation exposure. *Adv. Space Res.* **22**, 1631–1641 (1998).
110. Monarch, R. (munro). *Human-in-the-Loop Machine Learning: Active Learning and Annotation for Human-centered AI*. (Simon and Schuster, 2021).
111. Wang, A., Nguyen, D., Sridhar, A. R. & Gollakota, S. Using smart speakers to contactlessly monitor heart rhythms. *Commun Biol* **4**, 319 (2021).
112. Park, S.-M. *et al.* A mountable toilet system for personalized health monitoring via the analysis of excreta. *Nat Biomed Eng* **4**, 624–635 (2020).
113. Lassmann, M. & Eberlein, U. The Relevance of Dosimetry in Precision Medicine. *J. Nucl. Med.* **59**, 1494–1499 (2018).
114. Blue, R. S. *et al.* Limitations in predicting radiation-induced pharmaceutical instability during long-duration spaceflight. *NPJ Microgravity* **5**, 15 (2019).

115. NASA: Artemis. https://www.nasa.gov/specials/artemis/.
116. Hook, J. V. *et al.* Nebulae: A proposed concept of operation for deep space computing clouds. in *2020 IEEE Aerospace Conference* (IEEE, 2020). doi:10.1109/aero47225.2020.9172264.
117. HPC in Space: An Update on Spaceborne Computer after 1+ Year on the ISS. https://sc18.supercomputing.org/proceedings/bof/bof_pages/bof172.html.
118. McIntyre, A. B. R. *et al.* Single-molecule sequencing detection of N6-methyladenine in microbial reference materials. *Nat. Commun.* **10**, 579 (2019).
119. Azar, J., Makhoul, A., Barhamgi, M. & Couturier, R. An energy efficient IoT data compression approach for edge machine learning. *Future Gener. Comput. Syst.* **96**, 168–175 (2019).
120. Medical Operations: MRID Search. https://lsda.jsc.nasa.gov/MRID.
121. Goecks, J., Jalili, V., Heiser, L. M. & Gray, J. W. How Machine Learning Will Transform Biomedicine. *Cell* **181**, 92–101 (2020).
122. Banbury, C. R. *et al.* Benchmarking TinyML Systems: Challenges and Direction. *arXiv [cs.PF]* (2020).
123. Wang, Y., Yao, Q., Kwok, J. T. & Ni, L. M. Generalizing from a Few Examples: A Survey on Few-Shot Learning. *ACM Comput. Surv* **1**, (2020).
124. Marković, D., Mizrahi, A., Querlioz, D. & Grollier, J. Physics for neuromorphic computing. *Nature Reviews Physics* **2**, 499–510 (2020).
125. Liu, D., Yu, H. & Chai, Y. Low‑power computing with neuromorphic engineering. *Advanced Intelligent Systems* **3**, 2000150 (2021).
126. Hu, F., Xie, D. & Shen, S. On the Application of the Internet of Things in the Field of Medical and Health Care. in *2013 IEEE International Conference on Green Computing and Communications and IEEE Internet of Things and IEEE Cyber, Physical and Social Computing* 2053–2058 (2013).
127. Cheng, X. & Liu, H. A Novel Post-Processing Method Based on a Weighted Composite Filter for Enhancing Semantic Segmentation Results. *Sensors* **20**, (2020).
128. Jiang, H. & Nachum, O. Identifying and Correcting Label Bias in Machine Learning. *arXiv [cs.LG]* (2019).
129. Krueger, D. *et al.* Out-of-Distribution Generalization via Risk Extrapolation (REx). **139**, 5815–5826 (2021).
130. Tuan Nguyen, A., Tran, T., Gal, Y. & Baydin, A. G. Domain Invariant Representation Learning with Domain Density Transformations. *arXiv [cs.LG]* (2021).
131. Israel, D. J., Edwards, B. L. & Staren, J. W. Laser Communications Relay Demonstration (LCRD) update and the path towards optical relay operations. in *2017 IEEE Aerospace Conference* 1–6 (2017).
132. Adler-Milstein, J., Chen, J. H. & Dhaliwal, G. Next-Generation Artificial Intelligence for Diagnosis: From Predicting Diagnostic Labels to 'Wayfinding'. *JAMA* (2021) doi:10.1001/jama.2021.22396.
133. Erdemir, A. *et al.* Credible practice of modeling and simulation in healthcare: ten rules from a multidisciplinary perspective. *J. Transl. Med.* **18**, 369 (2020).
134. IEEE Standard Model Process for Addressing Ethical Concerns during System Design. *IEEE Std 7000-2021* 1–82 (2021).
135. Joly, Y., Saulnier, K. M., Osien, G. & Knoppers, B. M. The ethical framing of personalized medicine. *Curr. Opin. Allergy Clin. Immunol.* **14**, 404–408 (2014).
136. Li, T., Sahu, A. K., Talwalkar, A. & Smith, V. Federated Learning: Challenges, Methods, and Future Directions. *IEEE Signal Process. Mag.* **37**, 50–60 (2020).
137. Green, R. C., Lautenbach, D. & McGuire, A. L. GINA, genetic discrimination, and genomic medicine. *N. Engl. J. Med.* **372**, 397–399 (2015).
138. Lavin, A. *et al.* Technology Readiness Levels for Machine Learning Systems. *arXiv [cs.LG]* (2021).
139. Reynolds, R. J. & Shelhamer, M. Introductory Chapter: Research Methods for the Next 60 Years of
25

Space Exploration. in *Beyond LEO* (ed. Reynolds, R. J.) (IntechOpen, 2020).
140. Nangle, S. N. *et al.* The case for biotech on Mars. *Nat. Biotechnol.* **38**, 401–407 (2020).
141. Auñón-Chancellor, S. M., Pattarini, J. M., Moll, S. & Sargsyan, A. Venous Thrombosis during Spaceflight. *N. Engl. J. Med.* **382**, 89–90 (2020).
142. Marshall-Goebel, K. *et al.* Assessment of Jugular Venous Blood Flow Stasis and Thrombosis During Spaceflight. *JAMA Netw Open* **2**, e1915011 (2019).
143. Hanson, A. *et al.* A Model-Based Systems Engineering Approach to Exploration Medical System Development. in *2019 IEEE Aerospace Conference* 1–19 (2019).

Space Exploration. in *Beyond LEO* (ed. Reynolds, R. J.) (IntechOpen, 2020).
140. Nangle, S. N. *et al.* The case for biotech on Mars. *Nat. Biotechnol.* **38**, 401–407 (2020).
141. Auñón-Chancellor, S. M., Pattarini, J. M., Moll, S. & Sargsyan, A. Venous Thrombosis during Spaceflight. *N. Engl. J. Med.* **382**, 89–90 (2020).
142. Marshall-Goebel, K. *et al.* Assessment of Jugular Venous Blood Flow Stasis and Thrombosis During Spaceflight. *JAMA Netw Open* **2**, e1915011 (2019).
143. Hanson, A. *et al.* A Model-Based Systems Engineering Approach to Exploration Medical System Development. in *2019 IEEE Aerospace Conference* 1–19 (2019).

**Supplement 1 - Workshop Overview**

    To explore the future role of AI-modeling in space biology and health, NASA held a workshop in June 2021.  The workshop was organized by the NASA Space Biology Program within the Biological and Physical Science Division, part of the NASA Science Mission Directorate.  The NASA Human Research Program also supported and participated.  The workshop gathered a cohort of external-to-NASA AI-modeling subject matter experts (SME) in the fields of digital health, computer science, bioinformatics, medicine, microbiology, biomedical imaging and computational biology.

    The workshop's first day was organized to educate the AI-modeling SME cohort regarding: (1) long-term required biological and health capabilities needed for Lunar, Martian and deep space missions, (2) statuses of relevant data repositories, their content-structure and overall workflow of the current data resources to be mined-utilized, (3) current space-relevant biological AI, modeling and data science projects, (4) the unique statistical, data volume, cross-comparison and logistical challenges of data pertaining to astronaut health and space biological sciences.  Select 'central domain topics' guided the workshop:
• AI and Modeling for Knowledge Discovery: 'Omics and other Space Biological Data
• AI Applications in Imaging Space Biology Research Data (including Behavioral Analysis AI Tools for Space Data)
• Precision Medicine Utilization of AI
• Data Collection through Wearables, Sensors, Monitoring Hardware Systems and Integration with AI and Modeling Power
• Space Health Risk Predictions through AI, Modeling, Network Analyses
• Spaceflight Countermeasure Predictions Utilizing AI, Modeling, and Network Analyses
• AI Applications for Microbiology and Synthetic Biology
• AI Techniques and Translational Science Across Model Organisms and Species Toward Human Health

    On the second day of the workshop, the SME cohort and space-related researchers outlined AI and modeling recommendations and concepts for the next decade in space biology and space health.

# Supplement 2 - Workshop Flyer

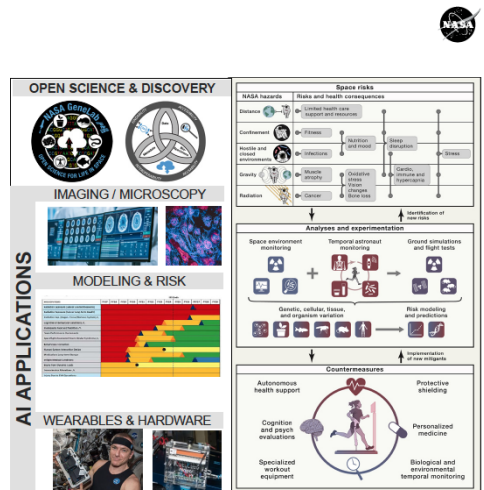

# Goals, Products, Format, Participants, Domain Topics

**Goals of the Workshop:**
- Identify AI technological needs for the next 10 years to address scientific gaps of space biology and health
- Develop strategic framework to fulfill gaps, emphasizing research consortiums, and funding mechanisms involving public and private partnerships
- Generate roadmap for AI and Modeling in space biology and health with key necessary capabilities and potential NASA campaigns

**Products:**
- White paper for NASEM-CBPSS and their Decadal Survey 2023-32, which workshop participants are encouraged to coauthor/cosign
- Pre-arranged peer-reviewed Journal Publication Summary

**Format:** Virtual workshop
**Length of workshop:** 2 days
- Day 1 – Overview, Keynotes, Learn from AI SMEs, State of Space Biology and Health Breakout Session 1, Writing
- Day 2 – Incubation Meeting of invited AI SMEs, Report from SME's Incubation Meeting, Report from Breakout Session 1, Breakout Session 2, Report from Breakout Session 2, Writing

**Participation:** 90 invitees. ~70 external AI-Data-Biology SMEs (both academic & industry). ~30 NASA internal (HQ, JSC, ARC, MSFC, GRC, etc.). All those interested in attending and contributing to the workshop are welcome.

## Central Domain Topics in our Workshop:
- AI and Modeling for Knowledge Discovery: 'Omics and other Space Biological Data
- AI Applications in Imaging Space Biology Research Data (including Behavioral Analysis AI Tools for Space Data)
- Precision Medicine Utilization of AI
- Data Collection through Wearables, Sensors, Monitoring Hardware Systems and Integration with AI and Modeling Power
- Space Health Risk Predictions through AI, Modeling, Network Analyses
- Spaceflight Countermeasure Predictions Utilizing AI, Modeling, and Network Analyses
- AI Applications for Microbiology and Synthetic Biology
- AI Techniques and Translational Science Across Model Organisms and Species Towards Human Health

# Vision of AI and Modeling in Our Field, Suggested References

**The Vision of AI and Modeling in Our Field of Space Biology:**
The era of artificial intelligence (AI) has opened new possibilities for the fields of biological sciences, physical sciences, and medicine. Because AI techniques can automate scientific and analytical processes while reducing the burden of required time and resource constraints, there are high expectations for AI-enabled knowledge discoveries. In the context of space biology and the interests of NASA, the steady growth of quantitative biological data such as DNA sequencing and omics (genomics, epigenetics, proteomics, …) is enabling AI applications for the prediction of disease risks, the identification of potential therapeutics, and the improvement of astronaut health in general. Also, the fusion of AI tools with the field of traditional biological modeling holds promise in quickening the pace of knowledge discovery. In addition, space biology ought to leverage the well-established usage of AI for image processing for knowledge extraction in diagnostic radiology, histopathology, immunohistochemistry, and cognitive-behavioral analysis from animal-based videos. AI will play a crucial role in the development of semi-autonomous health support for flight medical officers through health monitoring and integration with wearable sensors. In the near-future, assistance in medical diagnostics will be enabled by integrating state-of-the-art biotechnologies such as in situ genetic sequencing capability with health monitoring, for real-time personalized health-risk management. Lastly of note, by pushing AI technologies to their limit to address space biological and health-related challenges, myriad applications from this endeavor will be applicable on Earth and benefit terrestrial health substantially.

## Suggested Background References to Understand the Field:
- Cell Package: The Biology of Spaceflight
- Fundamental Biological Features of Spaceflight: Advancing the Field to Enable Deep-Space Exploration
- Evidence Reports from the NASA Human Research Program
- Research Methods for the Next 60 Years of Space Exploration
- FAIRness and Usability for Open-access Omics Data Systems
- NASA GeneLab: interfaces for the exploration of space omics data
- Policy Considerations for Precision Medicine in Human Spaceflight
- FDL, 2020, CRISP, Astronaut Health Technical Memo
- Integrating Spaceflight Human System Risk Research
- Knowledge Network Embedding of Transcriptomic Data from Spaceflown Mice Uncovers Signs and Symptoms Associated with Terrestrial Diseases
- From the bench to exploration medicine: NASA life sciences translational research for human exploration and habitation Missions
- Comprehensive Multi-omics Analysis Reveals Mitochondrial Stress as a Central Biological Hub for Spaceflight Impact
- FDL, 2019, Harnessing AI to support medical care in space, Technical Memo
- Considerations for Wearable Sensors to Monitor Physical Performance During Spaceflight Intravehicular Activities
- AI's role in deep space

---

## Confirmed Speakers

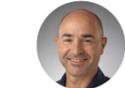
**Dr. Sylvain Costes**
Space Biosciences
Research Branch Chief (Acting)
GeneLab Project Manager
Lead of the Radiation Biophysics Laboratory
Senior Research Scientist, Code SCR
NASA Ames Research Center

**Dr. Lauren Sanders**
GeneLab Staff Scientist
NASA Ames Research Center

**Dr. Manil Maskey**
Senior Research Scientist,
Earth Science Data Systems
NASA Science Mission Directorate
NASA Headquarters

**Dr. Sergio Baranzini**
Department of Neurology,
Weill Institute for Neurosciences
University of California
San Francisco

**Dr. Jonathan Galazka**
GeneLab Project Scientist
NASA Ames Research Center

**Dr. Adrienne Hoarfrost**
NASA Postdoctoral Fellow
Rutgers University

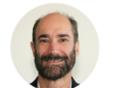
**Dr. Michael Snyder**
Stanford B. Ascherman
Professor and Chair,
Department of Genetics
Director, Stanford Center for Genomics and Personalized Medicine
School of Medicine,
Stanford University

**Dr. Kathleen Rubins**
NASA Astronaut and Molecular Biologist
NASA Johnson Space Center

**Dr. David Van Valen**
Assistant Professor of Biology and Biological Engineering
Caltech

**Dr. Greg Corrado**
Distinguished Scientist,
Google AI

**Dr. Patricia Parsons-Wingerter**
Biomedical Research Engineer
Low Gravity Exploration Technology (LTX)
NASA Glenn Research Center

**Dr. Mary Van Baalen**
Acting Chair, NASA Human Systems Risk Board
Space Medicine Operations Division,
NASA Johnson Space Center

**Dr. Robert Reynolds**
Baylor College of Medicine
Translational Research Institute for Space Health

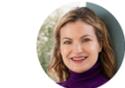
**Dr. Amina Qutub**
Associate Professor
Department of Biomedical Engineering
University of Texas, San Antonio
Director, UTSA-UT Health Graduate Group in Biomedical Engineering
Research Lead, AI MATRIX Consortium

**Dr. Casey Greene**
Professor, Department of Biochemistry and Molecular Genetics
Director of the Center for Health AI
University of Colorado
School of Medicine

**Dr. Sharmila Bhattacharya**
Program Scientist for NASA Space Biology
Biological and Physical Science Division
NASA Headquarters

**Dr. Jessica Keune**
Deputy Lead of NASA's Life Sciences Data Archive & Lifetime Surveillance of Astronaut Health
Space Medicine Operations Division,
NASA Johnson Space Center

**Dr. Aenor J Sawyer, MD, MS**
Director, UC Space Health
Director, UCSF Skeletal Health Service
Department Orthopaedic Surgery, UCSF
Co-Director, UCSF Center for Advanced 3D + Technologies

**Graham Mackintosh**
AI Projects Consultant
Bay Area Environmental Research Institute
NASA Advanced Supercomputer Division
NASA Ames Research Center, Code TN

## Agenda Day 1: Thursday, June 24 *all times PDT*

| Time | Session | Presenter / Panelists |
|---|---|---|
| 7:30 – 7:50 am | Welcome<br>Moving Space Biosciences to Knowledge-Based Platforms<br>Overview from 2021 Space Mission Directorate AI Workshop | Sylvain Costes NASA, BPS<br>Lauren Sanders NASA, BPS<br>Manil Maskey NASA SMD |
| 7:50 – 8:40 am | Keynote<br>AI Applications in Biosciences | Mike Snyder Stanford |
| 8:40 – 9:00 am | Break | |
| 9:00 – 10:00 am | Keynote<br>AI Applications in Space Biology | Kathleen Rubins NASA Astronaut |
| | **Learn From The Expert Mini-Symposium**<br>4 concurrent sessions | |
| 10:00 – 10:50 am | Single-cell biology in a Software 2.0 World | David Van Valen Caltech |
| | Frontiers of Machine Learning | Greg Corrado Google AI |
| | Artificial Intelligence, Systems Biology, Brain, and Health | Amina Qutub UT San Antonio |
| | AI in biology is powered by open data | Casey Greene University of Colorado |
| 10:50 – 11:10 am | Break | |
| | **Current State of Space Biology and Health – Resources and Challenges**<br>2 concurrent sessions | |
| 11:10 am – 1:00 pm | Space Biology and Physical Sciences Research | Sharmila Bhattacharya NASA Space Biology<br>Sergio Baranzini UCSF<br>Jon Galazka NASA, BPS<br>Adrienne Hoarfrost Rutgers University<br>Patricia Parsons-Wingarter NASA |
| | Human Health and Space Medicine | Mary Van Baalen NASA<br>Rob Reynolds Baylor<br>Jessica Keune NASA<br>Aenor Sawyer UCSF |
| 1:00 – 1:30 pm | Break | |
| 1:30 – 2:30 pm | **Breakout Session 1**<br>**Key Concept**<br>Knowledge and technology gaps in Central Domain Topics | |
| 2:30 – 2:50 pm | Break | |
| 2:50 – 4:00 pm<br>(Specific cohort of participants) | Breakout Session Report Writing | Organizers, moderators, and notetakers |

## Agenda Day 2: Friday, June 25 *all times PDT*

| Time | Session | Presenter / Panelists |
|---|---|---|
| 7:30 – 9:00 am<br>(Specific cohort of participants) | Incubation Meeting for Specific AI-Data Science SME Attendees<br>Identification of key needs and knowledge gaps<br>Development of vision for Decadal Survey white paper | Moderators<br>Amina Qutub UT San Antonio<br>Lauren Sanders NASA, BPS |
| 9:00 – 9:20 am | Break | |
| 9:20 – 10:40 am | Incubation Panel Report and Breakout Report | Amina Qutub UT San Antonio<br>Lauren Sanders NASA, BPS |
| 10:40 – 11:00 am | Break | |
| 11:00 – 11:10 am | Breakout Session Kickoff | Graham Mackintosh NASA |
| 11:10 am – 12:40 pm | **Breakout Session 2**<br>**Key Concept**<br>AI applications needed by space biology and health | |
| 12:40 – 1:10 pm | Break | |
| 1:10 – 1:40 pm | Open Discussion | Sylvain Costes NASA, BPS |
| 1:40 – 1:50 pm | **Instructions for Writing** | |
| 1:50 – 4:00 pm | **Breakout Writing Rooms (Optional)**<br>Central Domain Topic focus<br>Collation of content for Decadal Survey white paper and journal article | |

## Participants identify areas of expertise relating to "4 Quadrants" for breakout discussions

**4 Quadrants for All Participants:**
To organize discussions around AI, Modeling, space biology, medicine, the workshop organizers have developed a "4 Quadrants" framework to describe participant expertise. The areas of expertise within the quadrants are not meant to be all-inclusive, but rather to provide an overview of each quadrant. The organizers recognize that there are many ways to stratify the AI, modeling, biology, and medicine communities. The "4 Quadrants" framework was designed to focus discussion, disperse expertise across domains, and maximize the precious time attendees give to participation in the workshop.

**Purpose:**
The "4 Quadrants" framework will aid in distributing expertise evenly throughout the breakout discussions. These breakouts will have critical input to the white paper and journal paper deliverables.

**Actions:**
All participants will be asked to identify one or more main areas of expertise during registration, which will be used to stratify participants into quadrants.
Before the workshop, participants are asked to give some thought to examples of successes, challenges, and how to optimize the inherently inter-disciplinary and diverse teams required to leverage AI and Modeling towards space biology discovery and health.

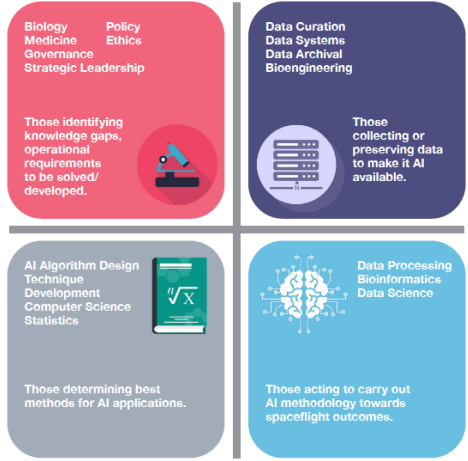

**Biology / Policy / Medicine / Ethics / Governance / Strategic Leadership** — Those identifying knowledge gaps, operational requirements to be solved/developed.

**Data Curation / Data Systems / Data Archival / Bioengineering** — Those collecting or preserving data to make it AI available.

**AI Algorithm Design / Technique Development / Computer Science / Statistics** — Those determining best methods for AI applications.

**Data Processing / Bioinformatics / Data Science** — Those acting to carry out AI methodology towards spaceflight outcomes.